\documentclass[aps,pra,showpacs,superscriptaddress,amssymb,twocolumn]{revtex4-1}
\usepackage{natbib}
\usepackage{graphicx}
\usepackage{enumitem}
\usepackage{appendix}
\usepackage{color}
\usepackage[export]{adjustbox}
\usepackage{epstopdf} 
\DeclareGraphicsExtensions{.pdf,.eps,.png,.jpg,.mps}
\usepackage[colorlinks=true,citecolor=blue,urlcolor=blue,linkcolor=blue]{hyperref}
\usepackage[position=top,singlelinecheck=off,justification=raggedright,caption=false]{subfig}
\usepackage{bm,amsmath}
\usepackage{dcolumn}
\usepackage{multirow}
\usepackage{hyperref}
\usepackage{float}
\input{Qcircuit}

\def\BEq{\begin{equation}}
\def\EEq{\end{equation}}
\def\BEqA{\begin{eqnarray}}
\def\EEqA{\end{eqnarray}}
\def\BW{\begin{widetext}}
\def\EW{\end{widetext}}
\usepackage{array}
\begin{document}
\title{Measurement-free implementations of small-scale surface codes for quantum dot qubits}

\author{H. Ekmel Ercan}
\author{Joydip Ghosh}
\author{Daniel Crow}
\author{Vickram N. Premakumar}
\author{Robert Joynt}
\author{Mark Friesen}
\author{S. N. Coppersmith}
\affiliation{Department of Physics, University of Wisconsin-Madison, Madison, Wisconsin 53706, USA}

\date{\today}

\begin{abstract}

The performance of quantum error correction schemes depends sensitively on the physical realizations of the qubits and the implementations of various operations.  For example, in quantum dot spin qubits, readout is typically much slower than gate operations, and conventional surface code implementations that rely heavily on syndrome measurements could therefore be challenging.  However, fast and accurate reset of quantum dot qubits--without readout--can be achieved via tunneling to a reservoir. Here, we propose small-scale surface code implementations for which syndrome measurements are replaced by a combination of Toffoli gates and qubit reset.  For quantum dot qubits, this enables much faster error correction than measurement-based schemes, but requires additional ancilla qubits and non-nearest-neighbor interactions.  We have performed numerical simulations of two different coding schemes, obtaining error thresholds on the orders of $10^{-2}$ for a 1D architecture that only corrects bit-flip errors, and $10^{-4}$ for a 2D architecture that corrects bit- and phase-flip errors.

\end{abstract}    

\maketitle

\section{Introduction}

Protecting quantum information against noise is one of the most important challenges for building a quantum computer \cite{LidarBrun}. Quantum error correction (QEC) addresses this issue by making use of redundancy \cite{Shor1995}. Among many QEC approaches, surface codes are considered to be particularly promising, with threshold error rates up to 1\% \cite{BravyiKitaev,Raussendorf,Fowler2012}. However, the practicality of a QEC procedure also depends on the physical implementation of the qubits.  Surface codes, as well as other standard QEC approaches, rely heavily on syndrome measurements, and the threshold calculations in the literature commonly assume that qubit measurements can be done within a gate time with failure probability equal to the gate failure probability. This assumption does not hold for several qubit implementations, including quantum dot qubits, for which readout times are several orders of magnitude longer than the gate times \cite{Petta2005,Kawakami2016,Scarlino2016,Brandur}. Therefore, conventional implementation of surface codes on these qubits seems challenging. On the other hand, DiVincenzo and Aliferis \cite{DiVincenzoAliferis} have shown that in a truly large scale quantum computer, error correction can be performed using slow measurements, because the measurements can take place concurrently within the many levels of concatenation required to achieve fault tolerance.  While of great interest for future applications, this result does not address the challenges faced by intermediate-scale implementations of 10-100 qubits, which realistically comprise only 1-2 levels of concatenation.

\begin{figure}
\includegraphics[width=0.45 \textwidth]{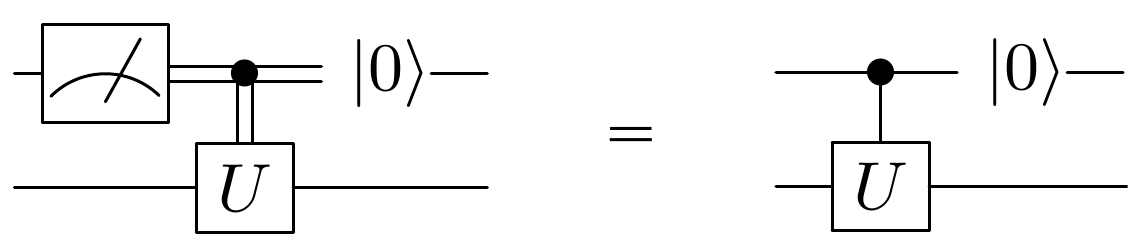}

\caption{A unitary operation determined by the result of a measurement of an ancilla followed by re-initialization of the ancilla (left circuit) is equivalent to a unitary controlled operation followed by the re-initialization of the control qubit (right circuit). Double lines in the left circuit represent a single classical bit. In both circuits, $\ket{0}$ indicates a rapid re-initialization of the qubit to its 0 state. For quantum dots, this involves tunneling to and from a reservoir.}
\label{fig:small_circuits}
\end{figure}

To overcome problems associated with measurement, it has been noted that the conventional widget used in syndrome measurements (a measurement followed by classical feedback), may be replaced by an alternative widget (a unitary gate operation followed by qubit re-initialization) \cite{NC,D19,Schindler,D21,Reed,D22}.  An example of such a measurement-free circuit is shown in Fig.~\ref{fig:small_circuits}. For many years it was thought that the error thresholds achievable using this strategy would be prohibitively low.  However, Paz-Silva \emph{et al.}\ demonstrated that measurement-free error correction for the Bacon-Shor code could have thresholds only about an order of magnitude worse than conventional schemes \cite{PazSilva}.   More recently, Crow \emph{et al.}\ improved these results by using redundant syndrome extractions and reported thresholds for three qubit bit-flip (BF), Bacon-Shor, and Steane codes that are comparable to measurement-based values \cite{Danny}. However, measurement-free surface code implementations have not yet been thoroughly investigated. Indeed, it has been argued that in this case, measurement-free implementations would significantly reduce the advantage of surface codes because the nontrivial classical matching problem used in this procedure relies on the results of error syndrome measurements to determine the recovery operations \cite{Fujii}.

Here, we show that this is not the case, for low-distance surface codes where the matching process can be simplified significantly \cite{PazSilva}. We investigate the performance of measurement-free QEC by simulating quantum circuits, obtaining threshold values for BF and surface codes of $2.0\times10^{-2}$ and $1.3\times10^{-4}$, respectively. These measurement-free thresholds can be compared to thresholds of $2 \times 10^{-2}$ and $8.0 \times 10^{-4}$ for the corresponding measurement-based procedures \cite{cheng,Tomita}.

This paper is organized as follows: In Sec.\ \ref{sec:stb_codes} we give an overview of QEC with surface codes and define necessary terms to make our discussion self-contained. In Sec.\ \ref{sec:mfree_ec} we explain in detail the changes we make to the conventional surface code implementations to  replace measurements with unitary operations and re-initializations. We first discuss the assumptions we make for quantum dot implementations in Sec.\ \ref{subsec:assump}, then demonstrate our method for a surface code restricted to one dimension, which is equivalent to the BF code in Sec.\ \ref{subsec:bf}, and generalize it to a full surface code in Sec.\ \ref{subsec:sc}. In Sec.\ \ref{sec:algo} we discuss the details of the error model and the simulation method that we use. We summarize our results in Sec.\ \ref{sec:results} and discuss them in Sec.\ \ref{sec:discussion}. 

\section{\label{sec:stb_codes}Stabilizer Codes}

The codes we consider in this paper belong to an important class of QEC codes known as \textit{stabilizer codes} \cite{NC}. The logical space of a stabilizer code is determined by a group of \textit{stabilizer operators} that leave this space unchanged under their action. Conventionally, stabilizers are repeatedly measured to ensure that the system remains in a simultaneous eigenstate of all of them. Over time, changes in the measurement outcomes indicate that errors have occured in the form of $X$, $Y$ or $Z$ qubit operators and that anti-commute with certain stabilizers. These operators are defined in terms of Pauli operators as $X=\sigma_x$, $Y=-i\sigma_y$ and $Z=\sigma_z$.

\begin{figure*}
\includegraphics[width=0.9 \textwidth]{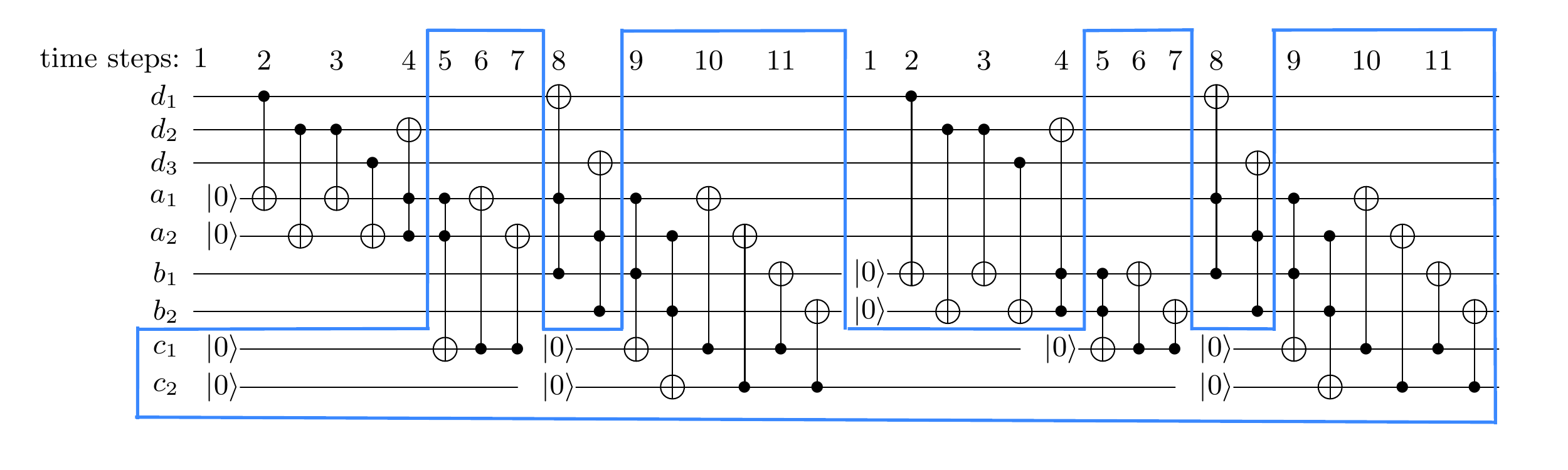}

\caption{The error-extraction and correction circuit for the bit-flip (BF) code, a one-dimensional surface code that only corrects bit-flip errors.  Here, $d_1$, $d_2$ and $d_3$ are data qubits and the rest are ancillas. The error information is extracted to ancilla pairs \{$a_1$,$a_2$\} and \{$b_1$,$b_2$\} in alternate time cycles, as shown in the figure. (A given cycle corresponds to time steps 1-11, where a particular step may be comprised of 1-2 gates that can be implemented simultaneously.) This two-cycle sequence ensures that single-data-qubit errors are always signaled by exactly two syndromes. A data error on $d_2$ flips both ancilla qubits that interact with $d_2$ in that cycle. A data error on $d_1$ or $d_3$ flips the pair \{$a_1$,$b_1$\} or \{$a_2$,$b_2$\}, respectively. Using these ancilla pairs as the controls and the relevant data qubits as the targets of the Toffoli gates, all single bit-flip errors are corrected. Whenever a Toffoli gate corrects an error, the control qubits must return to state $\ket{0}$ so that the matching is perfect, i.e. any syndrome is associated with at most one error. To do that, for each error-correcting Toffoli gate we employ another one targeting either $c_1$ or $c_2$ ancilla qubits and use these ancillas as the control qubits for two CNOT gates whose targets are the two control qubits of the error-correcting Toffoli gate. The portion of the circuit responsible for perfect matching is enclosed in the blue box in the figure. Excluding this subcircuit simplifies the QEC procedure considerably without significantly affecting its ability to correct single errors, as explained in the Appendix.}
\label{fig:1dcircuit}
\end{figure*}

To take an example, we first consider the logical space of the three-qubit BF code, which is spanned by $\ket{000}$ and $\ket{111}$ multiqubit states. These states are stabilized by two independent operators, $Z_{1}Z_{2}$ and $Z_{2}Z_{3}$. Here, the subscripts refer to physical qubits. As shown in Table~\ref{tab:table1}, each single bit-flip error $X_{1}$, $X_{2}$ or $X_{3}$ results in a different measurement outcome, or \textit{error syndrome}. Our measurement-free circuit that replaces the conditional operations with a combination of unitary operations and re-initializations ~\cite{NC} is shown in Fig.~\ref{fig:1dcircuit}.

\begin{table}[b]
\caption{\label{tab:table1}%
Stabilizers and syndrome values for the three-qubit bit-flip (BF) code. The two independent stabilizers are given in the first column. Each of the three possible bit-flip errors ($X_{1}$,$X_{2}$ or $X_{3}$) corresponds to different syndrome, uniquely identifying the error.}

\begin{tabular}{ c c c c c }
Stabilizer & \multicolumn{4}{c}{Error syndrome} \\
\toprule
\textrm{} & \textrm{$I$} & \textrm{$X_1$} & \textrm{$X_2$} & \textrm{$X_3$}\\
\colrule
$Z_{1}Z_{2}$ & 0 & 1 & 1 & 0 \\
$Z_{2}Z_{3}$ & 0 & 0 & 1 & 1 \\
\botrule
\end{tabular}
\end{table}

In a second example, we consider surface codes, which also utilize stabilizers. Surface code logical qubits can be arbitrarily large; here we focus on the specific 17-qubit surface code \cite{Tomita} shown in Fig.~\ref{fig:fig2}, which we refer to as surface-17 code. This is a distance-three logical qubit with nine data qubits and eight ancilla qubits that correspond to the eight independent stabilizers. These stabilizers, and logical X and Z operators are listed in Table~\ref{tab:stab}. Eigenstates of the logical Z operator are defined as the logical $\ket{0}$ and $\ket{1}$ states. We note that the measurement-free implementations of surface-17 will employ more than 17 qubits but the code will still be referred to as surface-17 as the structure of data qubits and independent stabilizers do not change. In particular, the measurement-free implementation of surface-17 that employs 25 qubits should not be confused with surface-25, which refers to a different distance-three surface code that employs 13 data qubits and 12 ancillas corresponding to 12 independent stabilizers.

In a general surface code, error syndromes cannot be uniquely matched with actual errors in the system; for a given set of error syndromes, physical error must be determined probabilistically. In a measurement-based surface code, this is done by storing a full history of stabilizer measurement outcomes from each error correction cycle, and using a classical minimum weight perfect matching (MWPM) algorithm~\cite{Edmonds} to match these syndromes with the most probable (i.e., the minimum weight) errors \cite{Fowler2012}. It is important to note that it is not necessary to correct errors immediately after they are detected. This is because the recovery operations belong to a Pauli group which is closed under the action of the Clifford group elements that contain all the gates needed for error correction \cite{Knill,DiVincenzoAliferis}.

Here, motivated by our interest in implementing error correction in experimental systems with moderate numbers of qubits, we investigate the performance of small surface codes that are implemented in a measurement-free fashion. As pointed out in Ref.~\onlinecite{Tomita}, syndrome--error matching for a small scale, distance-three, surface code is substantially simpler than for larger codes when only a short history of syndromes is used; in this case, MWPM can be reduced to the application of few simple rules which we summarize below.

\begin{table}
\caption {Stabilizers and logical operators for the bit-flip (BF) and surface-17 codes. Standard operators for the BF code \cite{NC} and the surface-17 code of Ref. \cite{Tomita} are also given for completeness of presentation.}

\begin{tabular}{c c c c}
\\
\multicolumn{2}{c}{BF code} & \multicolumn{2}{c}{Surface-17} \\
\toprule
\multicolumn{2}{c}{\textrm{\textit{Z} stabilizers}} & \textrm{\textit{X} stabilizers} & \textrm{\textit{Z} stabilizers}\\
\colrule
\multicolumn{2}{c}{$Z_{1}Z_{2}$} & $X_{2}X_{3}$ & $Z_{1}Z_{4}$\\
\multicolumn{2}{c}{$Z_{2}Z_{3}$} & $X_{7}X_{8}$ & $Z_{6}Z_{9}$\\
\multicolumn{2}{c}{}& $X_{1}X_{2}X_{4}X_{5}$  & $Z_{2}Z_{3}Z_{5}Z_{6}$\\
\multicolumn{2}{c}{}& $X_{5}X_{6}X_{8}X_{9}$  & $Z_{4}Z_{5}Z_{7}Z_{8}$\\
\toprule
\textrm{Logical \textit{X}} & \textrm{Logical \textit{Z}} & \textrm{Logical \textit{X}} & \textrm{Logical \textit{Z}}\\
\colrule
$X_{1}X_{2}X_{3}$ & $Z_{1}Z_{2}Z_{3}$ & $X_{3}X_{5}X_{7}$ & $Z_{1}Z_{5}Z_{9}$\\
\botrule
\end{tabular}
	\label{tab:stab}

\end{table}

\section{\label{sec:mfree_ec}Measurement-free error correction}
\subsection{\label{subsec:assump}Assumptions for quantum dot implementations}
We have argued that the measurement-free QEC scheme is appropriate for quantum dot qubits due to the fact that measurements are relatively slow, and therefore expensive, while qubit re-initialization is relatively fast, and therefore cheap.  We make use of this fact by using re-initialization repeatedly, and by assuming that the speed of re-initialization is similar to a unitary gate operation.  

In this work, we also make several other assumptions appropriate for quantum dot qubits.  First, to make up for the fact that measurement is expensive, we will assume that adding additional ancilla qubits to our circuits is relatively cheap.  This is motivated by the fact that quantum dots are considered to be a scalable qubit technology. Second, we note that long-range couplings between quantum dots are possible, in principle.  For example, strong coupling between a double quantum dot and a microwave resonator has recently been demonstrated \cite{Petta2016}, suggesting that long-range couplings between qubits could be achieved in the near future.  Hence, we assume that two- and three-qubit gates can be implemented natively in our circuits (i.e., in a single time step), even when the qubits are not in close proximity. For example, three-qubit Toffoli and controlled-controlled-Z (CCZ) gates play an important role in our scheme because, as will be explained later, correction of a data qubit error depends on the state of two ancilla qubits.  For simplicity here, we further assume that Toffoli and CCZ gates can be implemented with the same error rate as other gates.

\subsection{\label{subsec:bf}Bit-flip (BF) code}
The three-qubit BF code can be seen as an isolated line of three data and two ancilla qubits in a surface code. Therefore, the strategies developed in this setting can be easily generalized to more complicated cases, such as the surface-17 code. In this work, we only apply our error correction methods to logical identity gate operation, not to the logical X and Z gate operations described in Table~\ref{tab:stab}.

Below, we describe an error-correction strategy that is fault tolerant against single-qubit BF errors, including errors on both the data qubits and the ancillas. The method requires verifying information redundancy in space, as well as over time as our aim is to implement the conventional surface code error-correction strategy without measurements \cite{Fowler2012}. The latter involves storing and comparing syndrome information over consecutive cycles.  The BF and surface codes considered here both require two time cycles to complete their correction sequences.

The full error-detection and correction circuit for two consecutive cycles of the BF code is shown in Fig.~\ref{fig:1dcircuit}.  Here, the logical qubit is comprised of three data qubits ($d_1$-$d_3$) and eight ancillas. As discussed in Sec.\ \ref{sec:stb_codes}, there are two independent stabilizers and the syndrome information corresponding to each of these stabilizers is extracted to two ancilla qubits ($a_1$, $a_2$). Since we consider two consecutive time cycles, a separate pair of ancillas ($b_1$, $b_2$) is required to store the second set of stabilizer information.  Note that in the beginning of the first cycle of the figure, $b_1$ and $b_2$ carry information from the previous time cycle, which is not shown here. This scheme is different than measurement-based strategies, for which the result of the syndrome measurement is simply stored in a classical memory. In each cycle, the four syndromes are compared. If a match occurs, it signifies an error, which is then corrected by applying Toffoli gates (e.g., at time steps 4 and 8 in the figure). An important point is that each error syndrome occurs in only one matched pair, which is why such a matching is called ``perfect''. To ensure that the matching signaled by our circuit is ``perfect,'' any syndrome information used to correct an error should be removed from the system before the beginning of the next time cycle. This is done with the help of two additional ancillas ($c_1$ and $c_2$). For each error-correcting Toffoli gate T (e.g., at time step 4) there is another Toffoli gate T$'$ (time step 5) that employs the same control qubits but has a $c$ ancilla as its target. If an error is corrected by T, the $c$ ancilla which is initially in state $\ket{0}$, is flipped by T$'$, indicating that an error has been corrected. Two additional CNOT gates (time steps 6 and 7) and a subsequent re-initialization of $c_1$ and $c_2$ use this information to remove syndrome information from the system, preventing further matching.

The key to the fault tolerance of the above scheme is that
no single bit-flip error in either the data or ancilla qubits can cause a logical error. Using Toffoli gates for error correction requires having signals from two ancilla qubits. Therefore, for errors to propagate from ancilla qubits to data qubits at least two ancilla qubits must be flipped. Although there are cases where a single error can propagate into the two ancilla qubits, they are reset before being used in a Toffoli gate that can affect the data qubits. For example, an error occuring on qubit $c_1$ after time step 5 of the first cycle propagates into the qubits $a_1$ and $a_2$ but these qubits are reset in time step 1 (assuming cyclic repetition) before they can affect $d_2$ in time step 4.

As indicated by the blue box in Fig.~\ref{fig:1dcircuit}, a large portion of the full circuit is comprised of the gates that ensure perfect matching. Although perfect matching is necessary to correct certain error sequences, the additional gate operations it imposes can also introduce additional errors into the system that suppress the error threshold.  It is therefore interesting to study an alternative QEC circuit with the perfect-matching components removed. We will discuss the effects of these two approaches (i.e., longer circuit with perfect matching vs. shorter circuit with imperfect matching) on the performance of the surface code error correction in Sec.\ \ref{sec:results} and explain them in more detail in the Appendix. 

\begin{figure}
\includegraphics[width=0.48 \textwidth]{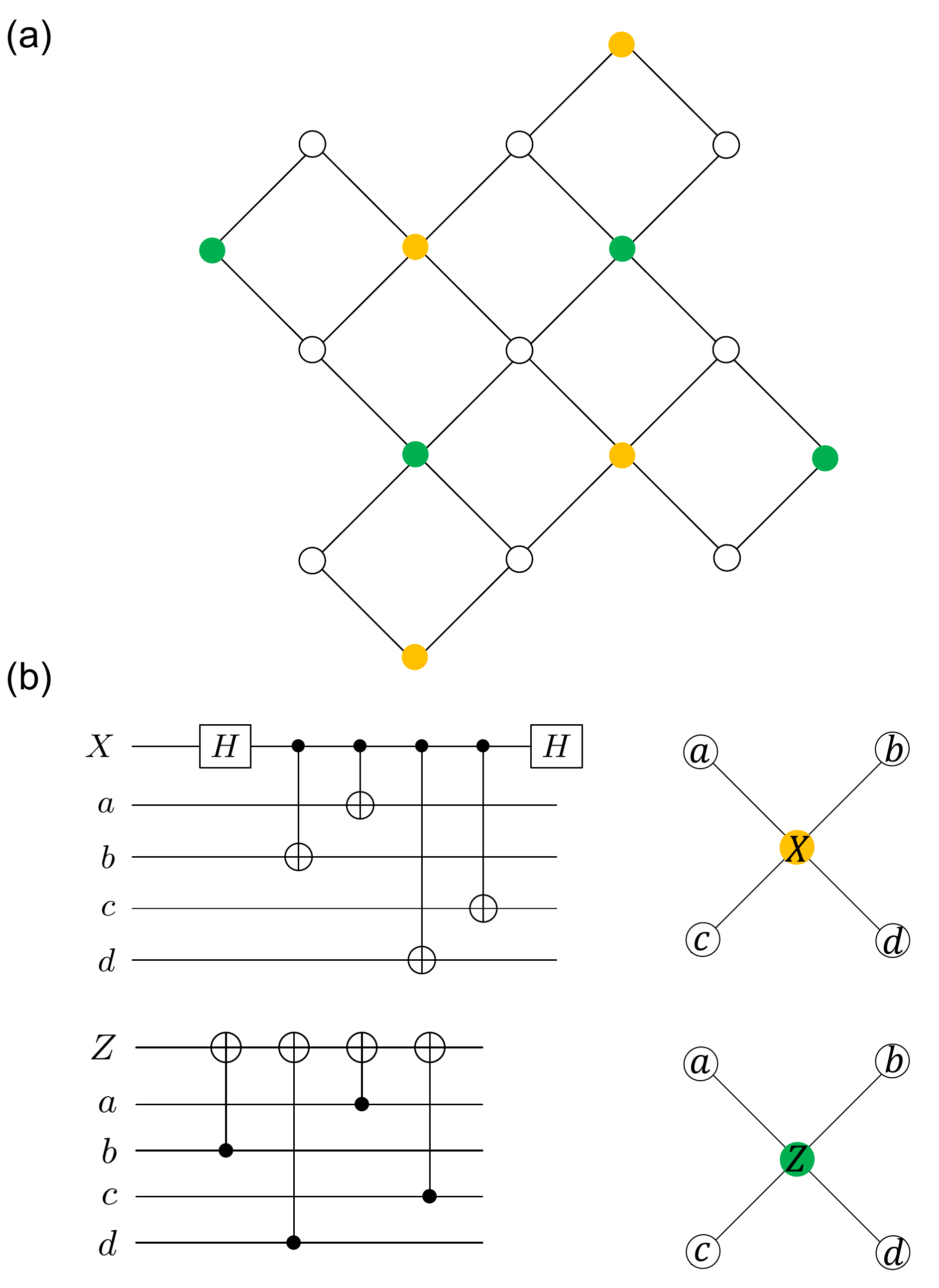}
      \caption{(a) Architecture of a 17-qubit, distance-three surface code, introduced in Ref.~\cite{Tomita}. Data qubits are denoted by the hollow circles whereas the X and Z syndrome qubits are denoted by orange (light) and green (dark) solid circles, respectively. (b) Standard syndrome-extraction circuits for X and Z syndromes, for the qubit orientations shown on the right. The circuits for the exterior ancillas that have only two neighboring data qubits involve fewer gates, with the same relative ordering.}
\label{fig:fig2}

\end{figure}

\begin{figure}
\includegraphics[width=0.4\textwidth]{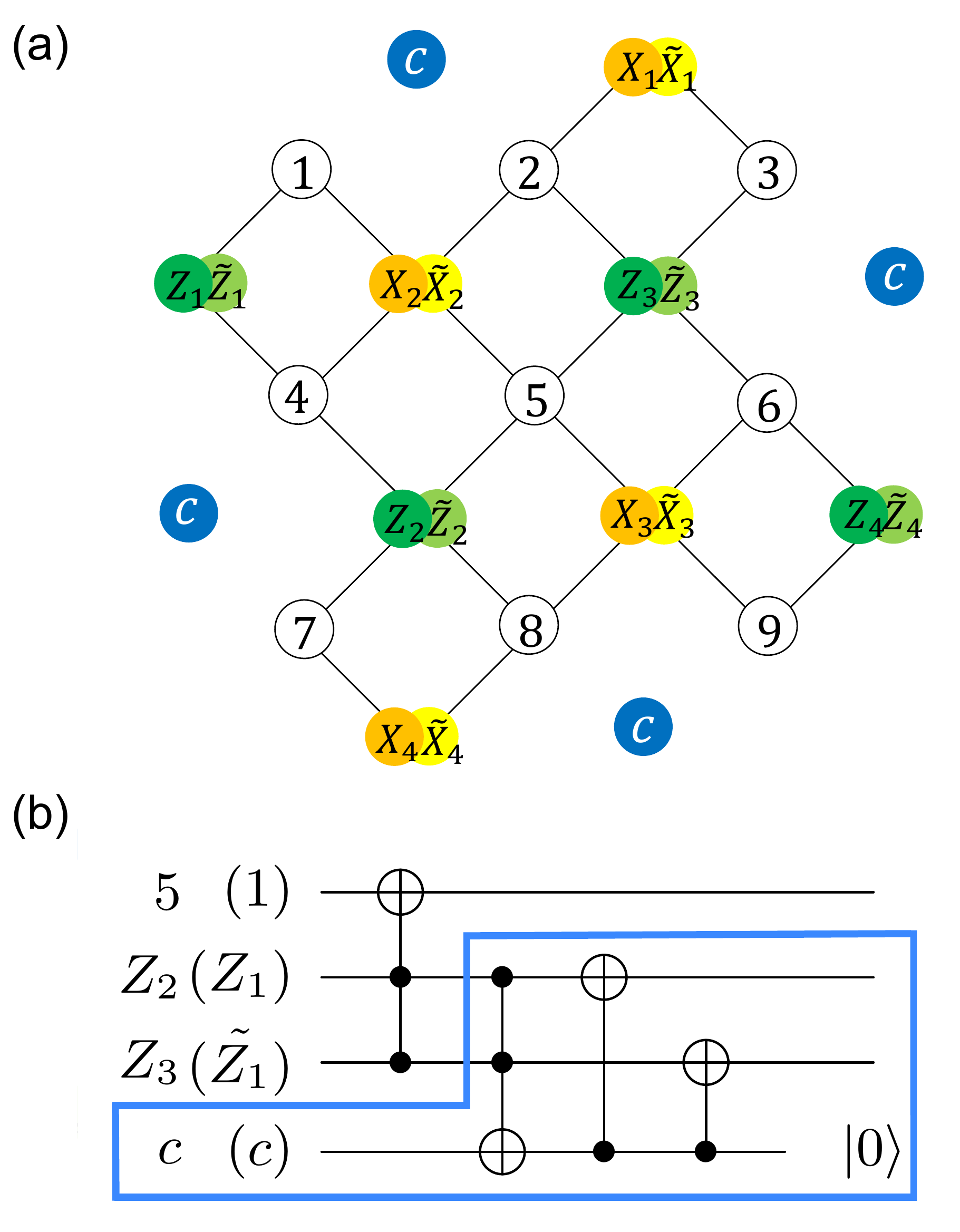}
      \caption{(a) 29-qubit system for implementing measurement-free error correction of a distance-three surface code for nine data qubits.  The data qubits are denoted as open circles labeled 1 to 9. These qubits and their error extraction syndromes, $ X_1,...,X_4,Z_1,...,Z_4 $ are present in the measurement-based architecture shown in Fig.~\protect\ref{fig:fig2}(a); twelve additional ancilla qubits (four c qubits and $ \tilde{X_1},...,\tilde{X_4},\tilde{Z_1},...,\tilde{Z_4}$) are present in the measurement-free architecture. Syndrome information is extracted into two sets of eight ancilla qubits, $ S_1=\{X_1,...,X_4,Z_1,...,Z_4\}$ and $ S_2=\{\tilde{X_1},...,\tilde{X_4},\tilde{Z_1},...,\tilde{Z_4}\}$, in alternate time cycles. Four ancilla qubits, denoted as blue filled circles labeled c, are used to remove used syndrome information from the system. The positions of the qubits in the figure do not necessarily represent the actual physical geometry, particularly since we have assumed long-distance couplings. (b) A portion of the circuit used to implement measurement-free $X$-error correction. Here, the syndrome has already been extracted and stored on the Z ancilla qubits. Two different cases are illustrated in the figure. Case 1: if there is an X error on qubit-5, a Toffoli gate controlled by two Z syndrome qubits from $S_1 $ corrects this error. All data qubits that have two neighboring ancilla qubits, of the same type (i.e., XX or ZZ), in the 17-qubit architecture are corrected in this way as well. Case 2 (in parentheses):  if there is an X error on qubit 1, the error-correcting Toffoli gate is controlled by one Z syndrome qubit from $ S_1 $ and one from $ S_2 $, corresponding to the syndromes extracted in two consecutive cycles. Other data qubits with only one neighboring ancilla qubit, of a given type, in the 17-qubit architecture are corrected in this way as well. The remaining gates remove the used syndrome information from the ancilla qubits with the help of the $c$ qubits. The Z error correction circuit is analogous, using X syndrome qubits instead of Z syndrome qubits and CCZ gates instead of Toffoli gates. As in Fig.~\ref{fig:1dcircuit}, the portion of the circuit in the box is responsible for perfect matching. }
\label{fig:fig3}

\end{figure}

\subsection{\label{subsec:sc}Distance-three surface code}
Here we generalize the strategy used for the BF code to the distance-three surface code of Refs.~\onlinecite{Tomita, Wootton}. A more thorough description of the surface code is presented in \cite{Fowler2012}. For our purposes, it is sufficient to note that isolated errors in the code (e.g., bit flips or phase flips) may be tolerated if they do not affect the topology of the encoded qubit.  However, a line of errors extending across the two-dimensional array does change the topology, and represents a logical error.  
Figure~\ref{fig:fig2}(a) shows the architecture of the surface-17 code appropriate for measurement-based error correction.  Here, the 9 data qubits indicated as open circles. Conventional syndrome extraction is performed on each proximal set of data qubits, as discussed in Ref. \cite{Tomita}, and the resulting syndrome values are stored in the 8 ancilla qubits indicated as colored circles in Fig.~\ref{fig:fig2}(b).  Two types of syndrome protocols (X and Z) are now required, since the two-dimensional code corrects both bit-flip and phase-flip errors.  The specific circuits needed for syndrome extraction involve CNOT and Hadamard gates, as shown in Fig.~\ref{fig:fig2}(b). 
To enable fault-tolerant, measurement-free error correction, we modify the 17-qubit architecture  along the same lines as was done for the BF code. Specifically, we add eight new ancillas that store the syndrome information from the preceding error correction cycle. In Fig.~\ref{fig:fig3}(a), the syndromes obtained in different time cycles are indicated as pairs of filled circles with different labels. Errors are then detected and corrected according to the following rules:
\begin{itemize}
\item If a X(Z) type data error occurs and the erroneous data qubit has two neighboring Z(X)-syndrome qubits in the original architecture in Fig.~\ref{fig:fig2}(a), implementation of a Toffoli (CCZ) gate controlled by those ancillas corrects the error.
\item If a X(Z) type data error occurs and the erroneous data qubit has only one neighboring Z(X)-syndrome qubit in the original architecture in Fig.~\ref{fig:fig2}(a), then error correction is performed only after the same syndrome is extracted again in the next time cycle. The correction is performed by applying a Toffoli (CCZ) gate controlled by two ancillas storing the syndrome information corresponding to the same stabilizer in two consecutive cycles. 
\end{itemize}

It is crucial to take into account the history of the syndromes for achieving fault-tolerance. To see this, consider an X error that occurs on qubit 5 in Fig.~\ref{fig:fig3}(a). Ideally, this causes the two neighboring Z syndrome qubits to get flipped by two CNOT gates in the syndrome extraction circuit [Fig.~\ref{fig:fig2}(b)]. However, if the error occurs between the application of these two CNOT gates, only one of the neighboring Z syndrome qubits will be flipped, falsely indicating an error on qubit 3. If the error-correction is done based on only this information (i.e., without comparing information from subsequent time cycles), an X gate will be applied to qubit 3 to``correct" a non-existing error, which adds a second error to the system. Now, the errors on qubits 3 and 5 cause qubit-7 to flip on a subsequent time cycle, resulting in a logical error. Obviously, such a scheme is not fault-tolerant as a single physical error can result in a logical error. On the other hand, in our circuit no error-correction is performed based on a single syndrome, as evident by the fact that all error-correcting gates (Toffoli or CCZ) involve three qubits. Using the circuit of Fig.~\ref{fig:fig3}(b), the error considered above would be ignored in the cycle in which it occurs, and it would get corrected in the following cycle, when both of the two neighboring Z syndrome qubits are flipped.

Once an error is corrected the syndrome information must be removed from the affected ancillas, as was done for the BF code. To do this we introduce four more ancillas, in contrast to the two ancillas needed for the BF code, to allow parallel operations to reduce the circuit depth. In total, this implementation employs 29 qubits. Figure~\ref{fig:fig3}(b) shows the portion of the circuit that corrects X errors on one of the data qubits (1 or 5) based on the extracted syndrome in $Z_1$, $Z_2$, $Z_3$, and $\tilde{Z_3}$.  To implement perfect matching, the syndrome information is then removed with the help of ancilla qubit c and the gates in the blue box. Using this scheme, a single error on syndrome qubits cannot affect the data qubits, which is important for achieving fault-tolerance. Fault-tolerance of the detection portion of the circuit is discussed in \cite{Tomita}. For the correction portion of the circuit, the discussion in the previous section also applies to this case, as the correction method is the same.

As noted in Sec.\ \ref{sec:mfree_ec}B, the extra gates required by the perfect-matching procedure may suppress the resulting error threshold.  By removing the gates in the blue box of Fig.~\ref{fig:fig3}(b), the number of qubits required for implementing the surface code is reduced from 29 to 25.  In Sec.\ \ref{sec:results} and the Appendix, we perform simulations to investigate both the full and reduced surface-code circuits. 


\section{\label{sec:algo}Simulations of algorithm performance}
We have checked the performance of the error-correcting algorithms by performing numerical simulations. These simulations use the algorithm developed by Aaronson and Gottesman, which evolves stabilizers rather than the full state \cite{Aaronson}. Toffoli and CCZ gates are not in the Clifford group and cannot generally be simulated with this method. However, in our simulations the control qubits of these gates are always in either $\ket{0}$ or $\ket{1}$ states. Hence, they can be``measured" and the gates can be performed in a classical manner, as in \cite{Danny}.

Our error model is composed of four different types of errors:
\begin{itemize}
\item Memory errors: after each ideal application of a single-qubit gate (including the identity), perform a $\pi$ rotation about the X, Y or Z axis, with each occurring with probability $p/3$.
\item Two-qubit gate errors: after each ideal application of a two-qubit gate, perform one of the 15 non-trivial tensor products of X, Y, Z and I, each occurring with probability $p/15$.
\item Three-qubit gate errors:  after each ideal application of a three-qubit gate, perform one of the 63 non-trivial tensor products of X, Y, Z and I, each occurring with probability $p/63$.
\item Initialization errors: after an ideal initialization to $\ket{0}$, perform an X rotation with probability $p$.
\end{itemize}

We emphasize that all error processes are assumed to occur with equal probability, for the sake of simplicity.
Our method for determining the logical error rate is based on estimating the time-to-failure for the error-correction scheme, similar to Refs. \cite{wang2010, Tomita}. We initialize our system to the logical $\ket{0}$ state and then continue running the simulation with errors occurring at the physical error rate $p$, until the system arrives at the logical $\ket{1}$. To make our simulation more efficient, we make the following observation: especially for small values of $p$, the simulation involves many cycles in which no error occurs and the the state of the system remains the same. Instead of simulating the full error-correction circuit for each of these error-free cycles, we sample how many of them occur consecutively. To do that, we define $P$ as the probability that no error occurs in a given cycle. Hence, $$P=(1-p)^{N},$$ where $p$ is the physical error rate and $N$ is the number of error sites in a cycle. Hence the probability of having no errors in $m-1$ consecutive cycles followed by at least one error in the $m^{th}$ cycle is $$P^{m-1}(1-P).$$ From this, the cumulative probability that the last, and only the last, cycle contains error(s) among all cycles up to and including $m^{th}$ cycle, where $m$ goes from 1 to $n$, is calculated as $$\sum_{m=1}^{n} P^{m-1}(1-P)=1-P^n,$$ and the number of error-free cycles $n$ is sampled as $$n=\frac{\ln{(1-r)}}{\ln{P}},$$ where $r$ is a uniformly distributed random number in $[0,1]$. Each time a number $n$ is sampled at least one error is required to be added to the system. We first determine exactly how many errors should be added using the conditional probability of having $k$ errors given that there is at least one error $$q(k)=\frac{\binom{N}{k}p^k(1-p)^{N-k}}{1-(1-p)^N}.$$ Then we choose $k$ error sites randomly and apply errors to them. At the end of each cycle we check whether  a logical error has occurred, or and whether the system has returned to its initial error-free state. If there is a logical error the simulation stops. Alternatively, if the state is error-free, another number $n$ is sampled and the process described above is repeated.

We have checked the reliability of this method by checking that the results agree with those obtained by simulating the full evolution (including error-free cycles) of the BF error-correction circuit in Fig.~\ref{fig:1dcircuit}, which is computationally less demanding than the surface-17 error correction in Fig.~\ref{fig:fig3}.

\section{\label{sec:results}Results}
\begin{figure}
  
    \includegraphics[width=.5\textwidth]{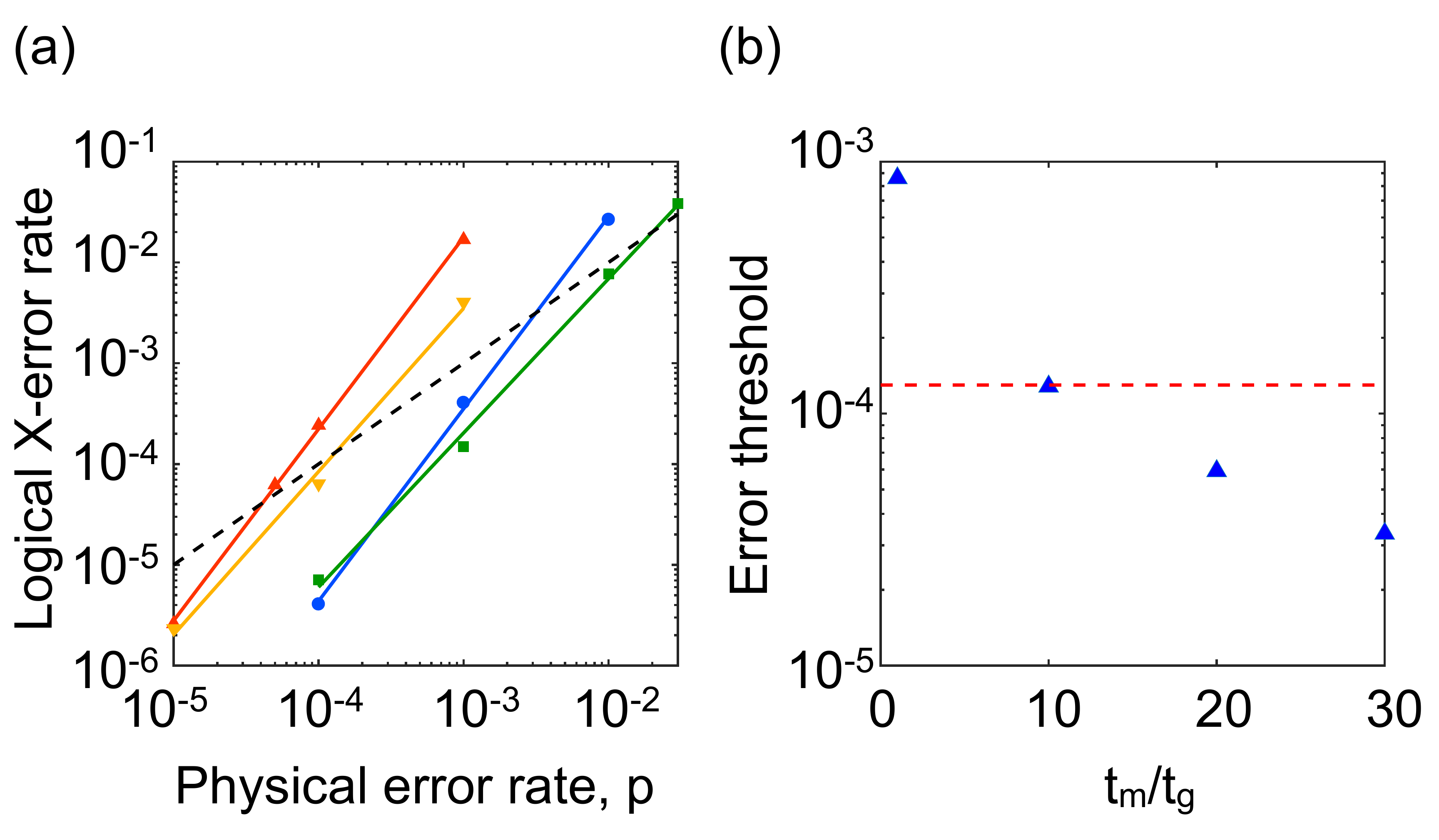}
     \caption{(a) Numerical results for the logical X error rates $p_\text{log}$ for the 1D BF code (blue and green lines), and the 2D surface-17 code, (red and orange lines) with measurement-free QEC and without error correction (black dashed line) as a function of the physical error rate $p$. 
The error threshold value is estimated as the value where the physical error rate is equal to the logical error rate. For the BF code, $p_\text{th,1D} \approx 3.2\times10^{-3}$, and for the 2D surface-17 code, $p_\text{th,2D} \approx 4.2\times10^{-5}.$ The simplified circuits for the BF and surface-17 codes, not including perfect matching (green and orange lines) produce higher thresholds of $p_\text{th,1D} \approx 2.0\times10^{-2}.$ and $p_\text{th,2D} \approx 1.3\times10^{-4}.$ (b) Measurement-based-threshold values, which were simulated here by using the circuit and look-up-table decoder in Ref. \cite{Tomita}, for surface-17 code at different measurement time/gate time ($t_m/t_g$) ratios (blue triangles) in comparison to the measurement-free threshold (red dashed line). Increasing measurement time reduces the error threshold as during the measurement of the ancilla qubits, the data qubits sit idle without protection against errors. When the measurement time vs. gate time ratio is around 10, measurement-free method produces a better threshold than the measurement-based method. }

\label{fig:plot}
\end{figure}

The numerical results for the logical-error rate $p_\text{log}$ as a function of the physical-error rate $p$ for the BF and surface-17 codes are shown in Fig.~\ref{fig:plot}(a).
The error threshold $p_\text{th}^\text{(X)}$ is defined as the point where these curves cross the dashed line  $p=p_\text{log}$ shown in Fig.~\ref{fig:plot}(a) \footnote{Note that different definitions for the error threshold exist in the literature. For instance, Crow \emph{et al.}\ \cite{Danny} define \textit{threshold} as we do here, whereas in Ref.\ \cite{Tomita} the term \textit{pseudothreshold} is used, reserving the term \textit{threshold} for the asymptotic threshold as the distance of the code goes to infinity.}. Based on these simulations, we estimate threshold values for the BF and surface-17 codes to be $3.2\times10^{-3}$ and $4.2\times10^{-5}$, respectively. If we remove the portions of the circuit used to implement perfect matching (i.e., the blue boxes in Fig.~\ref{fig:fig3} and in Fig.~\ref{fig:1dcircuit}) the thresholds for the BF and surface-17 codes improve to $2.0\times10^{-2}$ and $1.3\times10^{-4}$, respectively, which we attribute to the reduced number of error sites in the simpler scheme, and the fact the simplifications in the circuit do not affect its ability to correct single errors, as discussed in the Appendix. We note that the threshold values reported here correspond to logical X errors, which captures only two out of the three possible logical errors (X,Y and Z) that may occur. In Table~\ref{tab:table3}, we compare these values to measurement-based results for the same quantity, obtained elsewhere in the literature.

It is important to note that direct comparisons between different QEC schemes, such as those presented in Table~\ref{tab:table3}, can be complicated by the use of different simulation parameters, such as the logical coding schemes, the error models and the allowed gates. For instance, the measurement-based and measurement-free thresholds for the BF code are the same, however, the error model used in Ref. \cite{cheng} is slightly different model than the one used in this study.
\section{\label{sec:discussion}Discussion}

\begin{table}
\caption {\label{tab:table3}%
Comparison of measurement-free thresholds with measurement-based thresholds. The measurement-free BF and surface-17 codes studied here yield thresholds about an order of magnitude lower than their measurement-based counterparts, which we obtain from the literature. We note that the measurement-based threshold for surface-17 code, reported in Ref. \cite{Tomita}, was obtained using a look-up-table decoder that is based on a short history of syndromes. It is expected that the threshold obtained with a more sophisticated decoder would be higher \cite{fowler_surface25}.}

\centering
\begin{tabular}{  m{0.2\linewidth}  m{0.4\linewidth} m{0.4\linewidth}  } 
\\
\toprule

Code & Measurement-based threshold, $p_{\text{th}}^{(X)}$ & Measurement-free threshold, $p_{\text{th}}^{(X)}$ \\
\hline
\\
Bit-flip & $2.0 \times 10^{-2}$ \cite{cheng} & $2 \times 10^{-2}$ \\
Surface-17 & $8.0 \times 10^{-4}$ \cite{Tomita} & $1.3\times 10^{-4}$\\
\botrule
\end{tabular}
\end{table}

We have described a method of implementing low-distance surface codes on physical systems where measurement times are long but initialization times are short, such as semiconducting quantum dot qubits. In this method, we replace syndrome measurements by a combination of fast re-initialization and unitary gates.  We also assume that Toffoli (controlled-controlled-not) and CCZ (controlled-controlled-Z) gates can be implemented efficiently and that fast initialization into the $\ket{0}$ state is available. The method relies on the idea of storing and comparing syndrome information from two consecutive error-correction cycles. We have specifically considered an 11-qubit architecture for a bit-flip (BF) code, and 29 and 25-qubit architectures for a distance-three surface code that employ this method. We have calculated the error thresholds for all three of these schemes. A summary of our results along with the corresponding measurement-based thresholds for comparison is given in Table~\ref{tab:table3}. We note that the, measurement-based threshold for the surface-17 code in this table was calculated using a simplified decoder optimized for limited-memory implementations \cite{Tomita}. It may be possible to obtain a slightly better threshold for the measurement-based case by using a more advanced decoder. Indeed, this was found to be true for the surface-25 code \cite{fowler_surface25}.

Our results demonstrate that the error threshold for measurement-free error correction for the surface code is less than an order of magnitude lower than for measurement-based schemes. While achieving this threshold will be challenging in the laboratory, the results obtained here clearly become significant in the limit of long measurement times.  To demonstrate this, in Fig.~\ref{fig:plot}(b) we show that when the measurement-time is about 10 times the gate time, our method begins producing a higher threshold than the measurement-based method. Since the fastest quantum dot spin readout to date is about 100 ns \cite{Barthel_2010}, while qubit re-initialization can be as fast as 1 ns \cite{Petta2006}, measurement-free error correction schemes are clearly of current interest.

\begin{acknowledgments}
The simulations for this work were performed using the University of Wisconsin Center for High Throughput Computing. The authors thank Dan Bradley for his technical support, and M. A. Eriksson, M. Saffman and Yuan-Chi Yang for helpful discussions. The authors acknowledge support from the Vannevar Bush Faculty Fellowship program sponsored by the Basic Research Office of the Assistant Secretary of Defense for Research and Engineering and funded by the Office of Naval Research through grant N00014-15-1-0029.
\end{acknowledgments}

\appendix*

\section{A comparison of results with and without perfect matching}

\begin{figure}
  
    \includegraphics[width=0.55\textwidth]{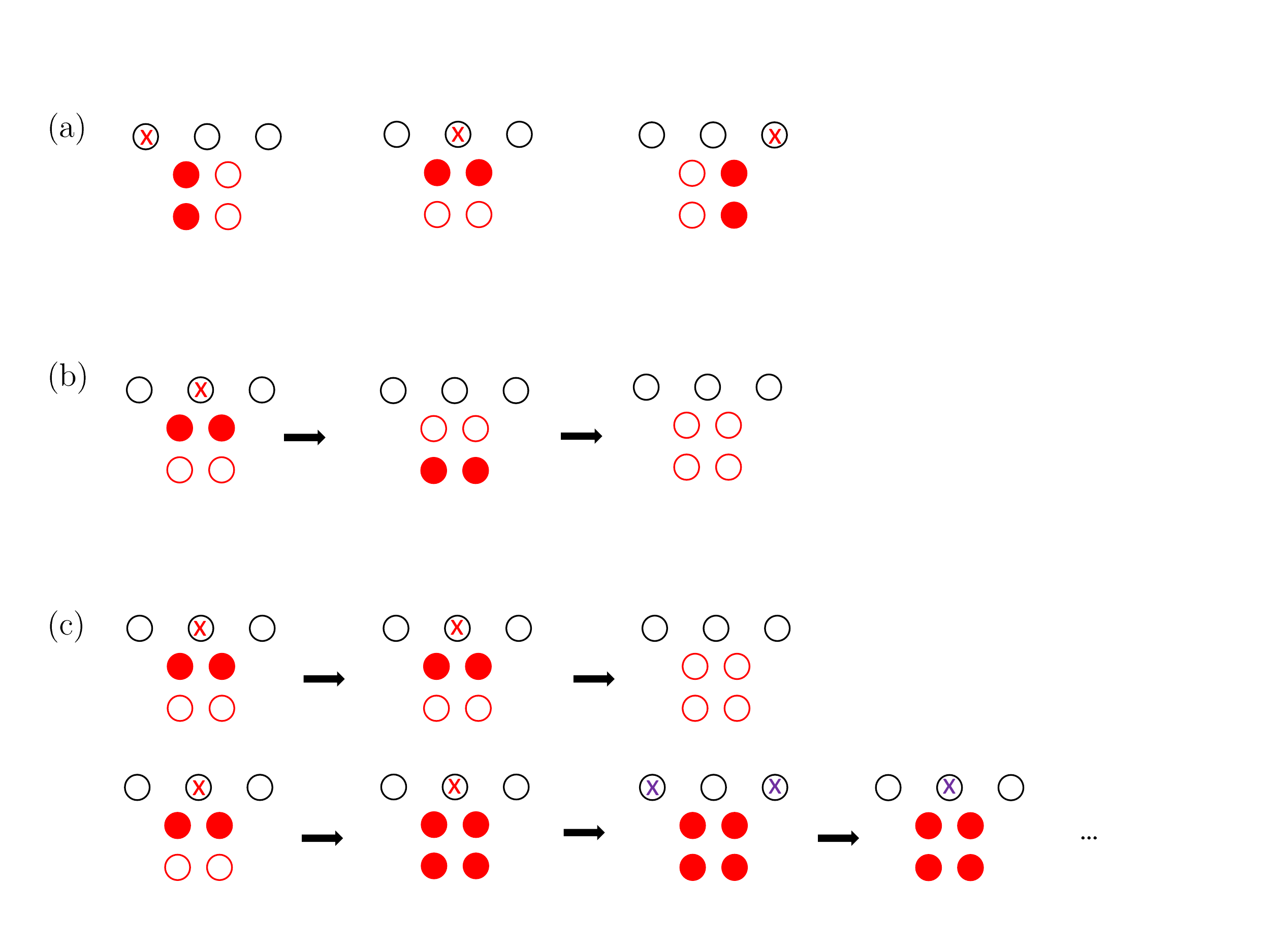}
     \caption{Examples showing the effect of removing used syndrome information removal in the BF code. The three (black) open circles in the top row represent data qubits. The four (red) circles below the data qubits represent the ancillas, where the middle row corresponds to the first time cycle, while the bottom row corresponds to the second time cycle. Flipped ancillas indicating a syndrome signal are filled and errors on data qubits are denoted with an X. (a) Here we show three different syndrome patterns that trigger different Toffoli gates to correct corresponding data errors. (b) In the case of a single data error, not removing used syndrome information is not harmful. For example, if there is an X error in the middle data qubit there will be two signals in the top row of ancillas, triggering a Toffoli gate that corrects the error. Even if these signals are not removed from the system, they will not trigger other Toffoli gates in the next cycle, and in the following cycle they will disappear completely. (c) We consider a case where errors appear in the middle data qubit in two consecutive time cycles. The first row shows that if the used information is removed both of these errors get corrected. In the second row, however, the information is not removed. This results in signals occurring in all the ancillas in the second cycle, which causes the Toffoli gates to correct errors in all three data qubits. One of these bit flips corrects the actual error, while the other two introduce new errors (shown in purple) resulting in signals in all ancillas in subsequent time cycles. The system then oscillates between having one and two errors indefinitely, making it vulnerable to errors that may occur in the future.  }
\label{fig:afig}
\end{figure}
In this section, we discuss the procedure for removing used syndrome information, which is done to avoid associating a syndrome with multiple error patterns. Although this step is crucial for achieving perfect matching in larger surface codes, in the main text we showed that excluding it actually improves our error threshold.  In Fig.~\ref{fig:afig}, where the three (black) open circles represent data qubits and the four (red) circles below them represent ancilla qubits, we show examples related to the BF code to illustrate that this procedure is only helpful for certain special cases where multiple errors occur, but are not well-separated in time. Eliminating the perfect-matching steps leaves the system unprotected against certain rare, multiple-error patterns but on the other hand decreases the number of error sites. Figure~\ref{fig:afig}(a) shows how we visualize the errors and error signals in this figure, when the circuit in Fig.~\ref{fig:1dcircuit} is implemented. Let the data qubits marked by a red X represent bit-flip errors and let the red filled ancilla qubits represent the error signals that control the error-correcting Toffoli gates. Here, the top row of the ancillas corresponds to results from the first time cycle, while the bottom row corresponds to results of the second time cycle. The three panels indicate that to correct the errors on the right-most and left-most data qubits, two error-correction cycles are needed whereas to correct an error on the middle data qubits only one is needed. In Fig.~\ref{fig:afig}(b), we consider a situation where removal of the used syndrome information does not benefit us. Specifically, we show three consecutive error-correction cycles, beginning with a single error on the middle data qubit. This error can be corrected in the first cycle as there are already two signals. Although there are still signals in the second cycle, this does not affect data qubits as these signals do not trigger any Toffoli gates. So, this particular error can be corrected without introducing extra errors into the system even when the used syndrome information is not removed. It can easily be seen that this is also the case for the other two possible single data errors. In Fig.~\ref{fig:afig}(c), we demonstrate an error pattern for which the removal of the used syndrome information is beneficial. In this case, errors occur on the middle data qubit in two consecutive cycles. in the first row, we consider the case where we remove the used syndrome information. Here, the first and second time cycles do not interfere with each other and both of the errors get corrected. In the second row, we consider the same errors in the case where we do not remove the used syndrome information. Here, the signals in the second cycle trigger all three Toffoli gates. While one of these gates corrects the actual error, the other two introduce new errors (indicated by purple X's) into the data qubits, causing both of the fresh ancillas to signal on a subsequent time cycle. It is easy to see that, in the absence of subsequent errors that would disturb the cycle, the system will now remain in a cycle where the four ancillas repeatedly signal errors and the data qubits oscillate between having one and two errors, failing to return to the error-free state.

\bibliography{qec_ekmel}
\end{document}